\documentclass[twocolumn,showpacs]{revtex4-1}
\usepackage{amsmath}
\usepackage{amssymb}
\usepackage{bm}
\usepackage{epsfig}
\usepackage{graphicx}
\usepackage{color}
\usepackage[T2A]{fontenc}
\usepackage[cp1251]{inputenc}
\usepackage[english]{babel}

\newcommand{\Sp}{\mathop{\rm Sp}\nolimits}

\begin{document}

\newcount\timehh  \newcount\timemm
\timehh=\time \divide\timehh by 60
\timemm=\time
\count255=\timehh\multiply\count255 by -60 \advance\timemm by \count255

\title{Spin fluctuations of non-equilibrium electrons and excitons in semiconductors}

\author{M.M. Glazov}

\affiliation{Ioffe Institute, 194021 St.-Petersburg, Russia}


\begin{abstract}
Effects related with deviations from thermodynamic equilibrium take a special place in the modern physics. Among those, non-equilibrium phenomena in quantum systems attract the highest interest. To date, the experimental technique of spin noise spectroscopy has became quite widespread, which makes possible to observe spin fluctuations of charge carriers in semiconductors both in equilibrium and non-equilibrium conditions. It calls for development of the theory of spin fluctuations of electrons and electron-hole complexes for non-equilibrium conditions. In this paper we consider a range of physical situations where a deviation from an equilibrium becomes pronounced in the spin noise. A general method of calculation of electron and exciton spin fluctuations in non-equilibrium state is proposed. A short review of theoretical and experimental results in this area is given.
\end{abstract}

\pacs{72.25.Rb,05.40.-a,74.25.nd}
\maketitle

\section{Introduction}\label{sec:intro}

Kinetics of non-equilibrium quantum systems is one of the most attractive areas of modern condensed matter physics. A deviation of an electronic system from the thermal equilibrium state caused by external static or alternating electromagnetic fields, injection of non-equilibrium charge carriers or optical pumping strikingly manifests itself in transport and optical phenomena in semiconductors and semiconductor nanostructures~\cite{Binder1995307,RevModPhys.74.895,RevModPhys.83.863,schnonequilibrium}. The diagram technique suggested by L.V. Keldysh~\cite{keldysh65} is an indispensable tool to study theoretically this sort of phenomena. 

Among a wide range of non-equilibrium effects one could mark out a special class, the phenomena related with fluctuations of observables under non-equilibrium conditions. By contrast to equilibrium fluctuations whose spectral power density is related with a linear susceptibility of the system by fluctuation-dissipation theorem~\cite{ll5_eng}, fluctuations at a non-equilibrium state can not be generally expressed via any response function~\cite{Lax1,springerlink:10.1007/BF02724353,ll10_eng,Blanter20001}. It leads to a great variety of fluctuation phenomena and, accordingly, necessitates to study individually each novel physical realization of a non-equilibrium noise. With a rise of diagram techniques suitable to study problems of physical kinetics~\cite{keldysh65,konst_perel_diag}, the graphical methods, including Keldysh technique, have been actively used  to study current fluctuations and generation-recombination noise of charge carriers~\cite{gancevich69:eng,gancevich70:eng,aronovivchenko71:eng,Kagan75:eng,PhysRevA.44.8072,altshuler1994nonequilibrium,PhysRevB.58.12883}.

\begin{figure}
\includegraphics[width=0.99\linewidth]{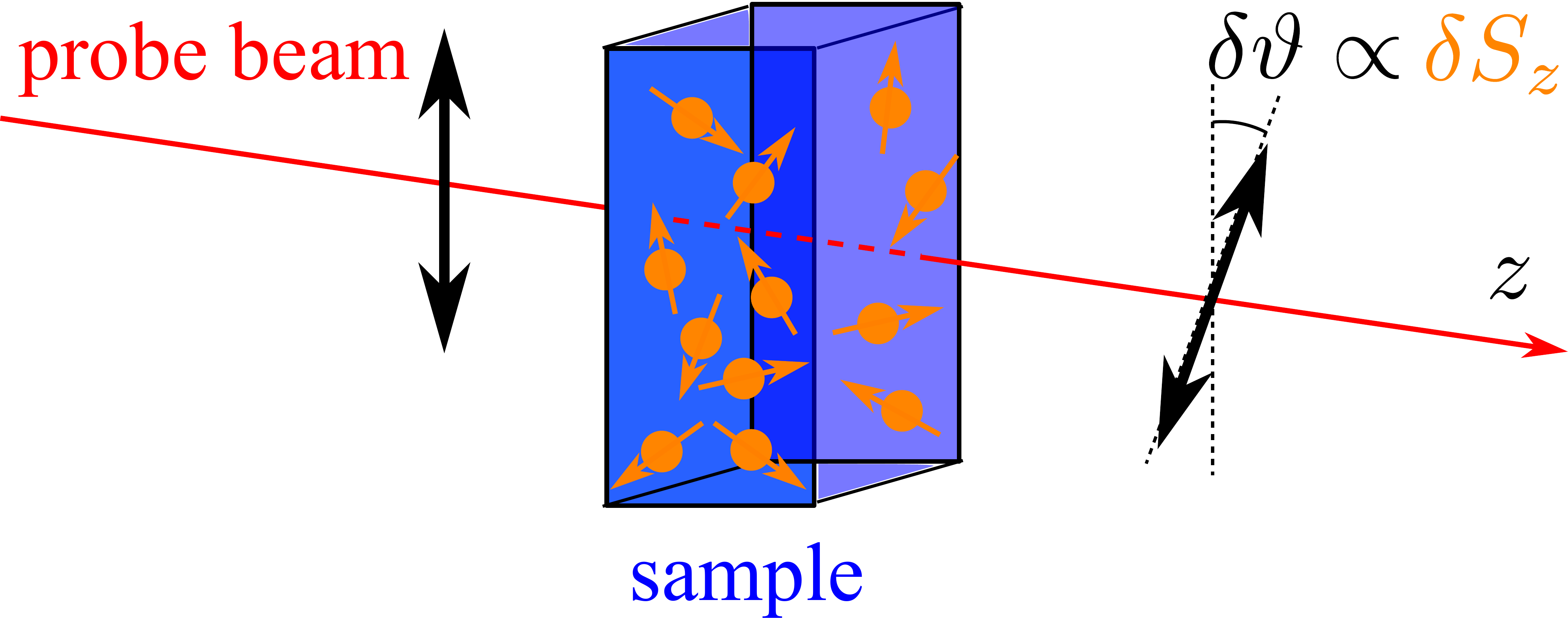} 
\caption{Schematic illustration of spin fluctuations detection: $z$ is the propagation axis of the linearly polarized probe beam, double arrows show the orientation of the linear polarization plane before and after the transmission through the sample.}\label{fig:SNS}
\end{figure}

The spin noise spectroscopy is being formed in the last decade as a novel line of investigations. The experimental method of spin noise observation has been first suggested and realized in the physics of atoms, it has been considered as an illustration of the fluctuation-dissipation theorem~\cite{aleksandrov81}. The linearly polarized ``probe'' light (whose frequency corresponds, as a rule, to the transparency or weak absorption region of a system) is transmitted through a sample. Fluctuations of polarization plane rotation $\delta \vartheta$ are detected either in transmission  (Faraday rotation fluctuations) or reflection (Kerr rotation fluctuations) geometry. Fluctuations of the rotation angle $\delta \vartheta$ are linearly related with the magnetization fluctuations in the system. Therefore, rotation angles autocorrelation function is directly proportional to the autocorrelation function of spins in the system:
\begin{equation}
\label{sns}
\langle \delta \vartheta(t') \delta \vartheta(t)\rangle \propto \langle \delta S_z(t') \delta S_z(t)\rangle,
\end{equation}
where $z$ is the light propagation axis, $S_z$ is the spin-$z$ component. Spin fluctuations have \emph{per se} quantum-mechanical nature, which stipulates considerable interest to their research. Moreover, development of spin fluctuations detection methods and applications of the spin noise spectroscopy to various systems~\cite{Crooker_Noise,Oestreich_noise,crooker2012,PhysRevB.89.081304} (for up-to-date review see Refs.~\cite{Zapasskii:13,Oestreich:rev}), possibilities to affect spin systems by external electromagnetic fields~\cite{PhysRevLett.113.156601}, a need to increase the method sensitivity by increasing the probe beam intensity~\cite{Glasenapp:2013fk,PhysRevB.89.081304}, make it topical to study spin fluctuations in semiconductors in non-equilibrium conditions.

The general formalism to calculate spin fluctuations in non-equilibrium systems is proposed in this paper. A brief overview of theoretical and experimental results on the non-equilibrium spin noise spectroscopy is given. Spin fluctuations in non-equilibrium electron systems are studied in Sec.~\ref{sec:sns:electrons}. Section~\ref{sec:sns:excitons} addresses the spin noise of neutral and charged excitons, two- and three-particle Coulomb-bound electron-hole complexes.

\section{Spin noise of non-equilibrium electrons}\label{sec:sns:electrons}

\subsection{General formalism}\label{sec:formalism}

Spin dynamics and fluctuations can be conveniently described in the density matrix formalism. Single-particle spin density matrix of free electrons in a semiconductor crystal has a form~\cite{dyakonov72,Glazov04_eeI}:
\begin{equation}
\label{dens}
\rho_{\bm k} = f_{\bm k} + \bm s_{\bm k} \cdot \bm \sigma.
\end{equation}
Here $\bm k$ is a quasiwavevector of an electron, $f_{\bm k}=\Sp\{\rho_{\bm k}/2\}$ is the spin-averaged occupation of the $\bm k$ state (electron distribution function), $\bm s_{\bm k} = \Sp\{\rho_{\bm k}\bm \sigma/2\}$ is the average spin in the given state (spin distribution function). We introduced $\bm \sigma = (\sigma_x,\sigma_y,\sigma_z)$, the pseudovector composed of Pauli matrices, unit $2\times 2$ matrix is omitted in Eq.~\eqref{dens}. It is known~\cite{springerlink:10.1007/BF02724353,ll10_eng} that microscopic distribution functions of particles and spins fluctuate around their mean values. These fluctuations  $\delta f_{\bm k}$, $\delta \bm s_{\bm k}$ are described by correlation functions  
\begin{multline}
\label{corr}
\langle \delta f_{\bm k}(t+\tau) \delta f_{\bm k'}(t)\rangle, \\ \langle \delta s_{\bm k,\alpha}(t+\tau) \delta s_{\bm k,\beta}(t)\rangle,~~\langle \delta f_{\bm k}(t+\tau) \delta s_{\bm k,\alpha}(t)\rangle.
\end{multline}
Here $\alpha,\beta=x,y,z$ are Cartesian subscripts. We consider spatially-homogeneous fluctuations in the stationary but, generally, non-equilibrium state, hence, the averaging in Eq.~\eqref{corr} is carried out over a time $t$ at the fixed $\tau$. One can take into account spatial inhomogeneity and non-stationarity following Refs.~\cite{Kagan75:eng,PhysRevA.44.8072,PhysRevB.58.12883}. Since, generally speaking, occupation number and spin operators taken at different moments of time do not commute, the observables are expressed via symmetrized combinations, e.g., $\langle \{\delta s_{\bm k,\alpha}(t+\tau) \delta s_{\bm k,\beta}(t) \}\rangle$, where $\{AB\} = (AB+BA)/2$. The curly brackets denoting symmetrization are omitted everywhere, unless it leads to a confusion.

Under standard conditions
\begin{equation}
\label{conditions}
\frac{\bar \varepsilon \tau}{\hbar} \gg 1, \quad \frac{\bar \varepsilon}{\hbar \omega} \gg 1,
\end{equation}
where $\bar \varepsilon$ is the mean kinetic energy of the electrons, $\tau$ is the characteristic relaxation time of the electron system, $\omega$ is the characteristic frequency of fluctuations, the single-particle density matrix~\eqref{dens} and correlation functions~\eqref{corr} satisfy kinetic equations. A most general way of its derivation is to use Keldysh diagram technique~\cite{keldysh65}. This method has been applied in the theory of non-equilibrium electron gas fluctuations in Ref.~\cite{Kagan75:eng} to analyse distribution function fluctuations, $\delta f_{\bm k}$, in a non-equilibrium Fermi gas and has made it possible to account for the dynamic screening of the Coulomb interaction. The method has been developed in Ref.~\cite{PhysRevB.58.12883} for semiconductors in the presence of optical excitation. A generalization of Keldysh technique has been suggested in Ref.~\cite{PhysRevA.44.8072}, which allows, in particular, to go beyond condition~\eqref{conditions}.

\begin{figure*}[hbt]
\includegraphics[width=0.99\textwidth]{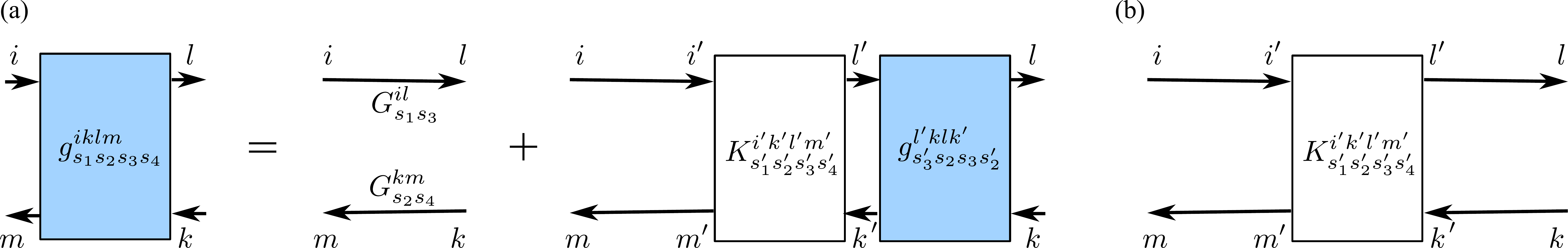} 
\caption{(a) Graphic representation of the correlation function, Eq.~\eqref{g:corr} (filled-in block). Empty block shows the kernel $K$.\\ (b) Block describing a source of correlation in Eq.~\eqref{same:t:dyn}.}\label{fig:g:corr:eq}
\end{figure*}

Below we briefly summarize the derivation of kinetic equations for spin correlation functions, Eq.~\eqref{corr}, in Keldysh technique following Refs.~\cite{Kagan75:eng,PhysRevB.58.12883}. We introduce the correlation function 
\begin{multline}
\label{g:corr}
g^{iklm}_{s_1s_2s_3s_4}(t+\tau+\delta', \bm k_1; t+ \delta, \bm k_2; t,\bm k_3;t+\tau, \bm k_4) = \\
\langle T_c a_{\bm k_1s_1}(t+\tau+\delta')_i  a_{\bm k_2s_2}(t+\delta)_k  a^\dag_{\bm k_3s_3}(t)_l  a^\dag_{\bm k_4s_4}(t+\tau+\delta')_m \rangle,
\end{multline} 
where $a_{\bm k,s}$ ($a_{\bm k,s}^\dag$) are the annihilation (creation) operators of an electron in the state $\bm k$ with the spin $s$ ($s_1, \ldots s_4 = \pm 1/2$), $T_c$ is the ordering operator at Keldysh contour, subscripts $i,k,l,m$ take two values ``$-$'' (1) and ``$+$'' (2), they enumerate upper and lower branches of the contour, $\delta,\delta'\to 0$. It is convenient to exclude from $g^{iklm}_{s_1s_2s_3s_4}$ a product of mean density matrices $\rho_{\bm k} \rho_{\bm k'}$ corresponding to the single-particle Greens functions $-G^{im}_{s_1s_4}G^{kl}_{s_2s_3}$~\footnote{There is a certain freedom in the choice of subscripts on the Keldysh contour~\cite{Kagan75:eng,PhysRevB.58.12883}, hence, it is convenient to put $i=+$, $k=-$, $l=-$, $m=+$ in the calculations~\cite{altshuler1994nonequilibrium}.}. It is instructive to represent the equation for the correlator~\eqref{g:corr} graphically, Fig.~\ref{fig:g:corr:eq}(a), where the arrows denote single-particle Greens functions, the filled-in block denotes the required function $g^{iklm}_{s_1s_2s_3s_4}$, the  empty block is the kernel $K^{i'k'l'm'}_{s_1's_2's_3's_4'}$, accounting for the electron-impurity, electron-phonon and electron-electron interactions.

The standard procedure~\cite{Kagan75:eng,PhysRevB.58.12883} allows us to reduce the integral equation in Fig.~\ref{fig:g:corr:eq}(a), to a kinetic equation for the ``shortened'' correlation function $g^{iklm}_{s_1s_2s_3s_4}(t+\tau,\bm k; t, \bm k')=g^{iklm}_{s_1s_2s_3s_4}(t+\tau, \bm k;t,\bm k';t\bm k'; t+\tau, \bm k)$, which provides the statistics of distribution functions fluctuations $\delta f_{\bm k}$, $\delta \bm s_{\bm k}$ [cf. Eq.~(20) from Ref.~\cite{PhysRevB.58.12883}]:
\begin{widetext}
\begin{multline}
\label{kinetic:2}
\frac{\partial}{\partial \tau} g_{s_1s_2s_3s_4}(t+\tau,\bm k; t, \bm k') + \frac{i}{\hbar} \sum_{s'} \left[\mathcal H_{s_1s'}g_{s's_2s_3s_4} - g_{s_1s_2s_3s'} \mathcal H_{s's_4}  \right]\\  + e\bm E \frac{\partial }{\partial \bm k} g_{s_1s_2s_3s_4}(t+\tau,\bm k; t, \bm k')  +\mathcal Q_{\bm k, \bm k'} \{ g\}  =\left(\frac{\partial}{\partial \tau} + \mathcal I_{\bm k}\right) \{g\} = 0.
\end{multline}
\end{widetext} 
Here $\mathcal H_{ss'} \equiv \mathcal H_{ss'}(\bm k)$ are the matrix elements of electron spin Hamiltonian $\mathcal H$, which takes into account both spin interaction with external static (and low-frequency, $\hbar\omega \ll \bar \varepsilon$) fields and the spin-orbit coupling, $\bm E$ is the external electric field, $e<0$ is the electron charge, $\mathcal Q_{\bm k, \bm k'} \{ g; \rho\}$ is the collision integral, which depends on the stationary single-particle density matrix of the system. Keldysh subscripts $i\ldots m$ together with arguments of the function $g$ are omitted in Eq.~\eqref{kinetic:2}, unless it results in a confusion. The evolution operator $\mathcal I_{\bm k}$ is also introduced in the second equality of formula \eqref{kinetic:2}, the subscript $\bm k$ means that the operator acts on the single-electron variables: $\bm k$, $s_1$, $s_4$.  Note, that the single-particle density matrix $\rho_{\bm k}$ satisfies, in the steady state, the analogous equation:
\begin{equation}
\label{kinetic:1}
\frac{i}{\hbar}\left[\mathcal H, \rho_{\bm k}  \right] + e\bm E \frac{\partial }{\partial \bm k} \rho_{\bm k}(t) +\mathcal Q_{\bm k} \{ \rho\} =0,
\end{equation} 
with the collision integral $\mathcal Q_{\bm k}\{\rho\}$. 

Physical meaning of Eq.~\eqref{kinetic:2} can be easily understood making use of Onsager hypothesis: A stochastic fluctuation of the electron density matrix $\delta \rho_{\bm k}$ evolves in the same way, as if this fluctuation has been prepared at the initial moment $\tau=0$ by external forces~\cite{springerlink:10.1007/BF02724353}. Hence, at fixed  $\bm k'$, $t$ and two spin subscripts $s_2,s_3$, the correlator $g_{s_1s_2s_3s_4}(t+\tau,\bm k; t, \bm k')$ satisfies standard kinetic equation for the spin density matrix. The only difference is in the form of the collision integral $\mathcal Q_{\bm k, \bm k'}\{g;\rho\}$, which could be derived from the integral $\mathcal Q_{\bm k} \{ \rho\}$ in Eq.~\eqref{kinetic:1} by linearization in small fluctuations. Note, that the expression for $\mathcal Q_{\bm k}\{\rho\}$ accounting for electron-electron collisions is reported in Ref.~\cite{Glazov04_eeI}, the formula for the collision integral  $\mathcal Q_{\bm k, \bm k'}\{g;\rho\}$ can be derived from the general expressions of Ref.~\cite{Glazov04_eeI} by decomposition in fluctuations.

Equation~\eqref{kinetic:2} should be supplemented by an initial condition: a value of the correlation function $g_{s_1s_2s_3s_4}(t+\tau,\bm k; t, \bm k')$ at $\tau=0$. Similarly to the theory of fluctuations of spin-averaged distribution function~\cite{springerlink:10.1007/BF02724353,Kagan75:eng,PhysRevB.58.12883,ll10_eng}, let us separate in the single-time correlator the part describing mean squares of fluctuations in a given quantum state
\begin{multline}
\label{same:t}
g_{s_1s_2s_3s_4}(t,\bm k; t, \bm k') = \delta_{\bm k \bm k'}\rho_{\bm k, s_2s_4}[\delta_{s_1s_3} - \rho_{\bm k, s_1s_3}] \\ + \Phi_{s_1s_2s_3s_4}(\bm k, \bm k').
\end{multline}
Function $\Phi_{s_1s_2s_3s_4}(\bm k, \bm k')$ describes nontrivial simultaneous correlation of fluctuations in the system. Making use of the graphical equation in Fig.~\ref{fig:g:corr:eq}(a) we may show that this function, as a function of each triple of variables $\bm k$, $s_1$, $s_4$ and $\bm k'$, $s_2$, $s_3$, satisfies the inhomogeneous stationary kinetic equation with the operators  $\mathcal I_{\bm k}$ и $\mathcal I_{\bm k'}$, respectively:
\begin{equation}
\label{same:t:dyn}
\left(\mathcal I_{\bm k} + \mathcal I_{\bm k'} \right) \left\{\Phi_{s_1s_2s_3s_4}(\bm k, \bm k')\right\} = \mathcal  L_{s_1s_2s_3s_4}(\bm k, \bm k').
\end{equation}
Formally, the source of correlations $\mathcal  L_{s_1s_2s_3s_4}(\bm k, \bm k')$ results from differentiation over $t$ of the block composed of four Greens function and the kernel  $K$, shown in Fig.~\ref{fig:g:corr:eq}(b). Physically speaking, the function $\mathcal L$ describes simultaneous and correlated variation of the spin density matrices of the states $\bm k$, $\bm k'$. For example, it is well known~\cite{kogan_shulman1,gancevich69:eng,gancevich70:eng,springerlink:10.1007/BF02724353,ll10_eng}, that even in the nondegenerate gas ($f_{\bm k} \ll 1$) electron-electron collisions serve as a source of pair correlations of occupation numbers of the states $\bm k$, $\bm k'$. This result can be generalized for the spin-polarized electron gas. In the simplest possible case where electrons are polarized along the single axis $z$, the correlations appear in the spin-diagonal elements of the density matrix, namely:
\begin{multline}
\label{ee:corr}
\mathcal  L_{s_1s_2s_1s_2}(\bm k, \bm k') = \sum_{\bm p,s_1', \bm p's_2'} W(\bm k s_1, \bm k's_2; \bm ps_1',\bm p's_2') \\ \times (
\rho_{\bm p,s_1's_1'} \rho_{\bm p',s_2's_2'} - \rho_{\bm k,s_1s_1} \rho_{\bm k',s_2s_2}),
\end{multline}
where $W(\bm k s_1, \bm k's_2; \bm ps_1',\bm p's_2')$ is the probability of the Coulomb scattering of the electron pair from the states  $\bm p$, $\bm p'$ with spins $s_1'$, $s_2'$ to the states $\bm k, \bm k'$ with spins $s_1$, $s_2$. Derivation and analysis of  general expressions for the pair correlations source $\mathcal  L_{s_1s_2s_3s_4}(\bm k, \bm k')$ is beyond the scope of the present paper.

Equations~\eqref{kinetic:2} and \eqref{same:t:dyn} make it possible to calculate amplitudes and dynamics of the electron gas spin fluctuations. For generality, we present the expression for the spin noise power spectra,
\begin{equation}
\label{fourier}
(\delta s_{\bm k,\alpha} \delta s_{\bm k',\beta})_\omega = \int_{-\infty}^\infty \langle \delta s_{\bm k,\alpha}(t+\tau) \delta s_{\bm k',\beta}(t) \rangle e^{\mathrm i \omega\tau} d\tau,
\end{equation}
because such kind of quantities~\footnote{As a rule, these quantities summed by $\bm k$ with some weight functions are observed in experiment.} can be directly measured in experiments~\cite{aleksandrov81,Crooker_Noise,Oestreich:rev,Glazov:15}.  The method outlined here can be easily extended for the case of localized electrons, where the electron-electron interaction is not important, while the hyperfine coupling of electron and host lattice nuclear spins plays a role. This method can be also used to calculate valley polarization and coherence fluctuations in multivalley semiconductors and nanosystems~\cite{Bareikis,PhysRevLett.113.046602}.

\begin{figure*}
\includegraphics[width=0.99\textwidth]{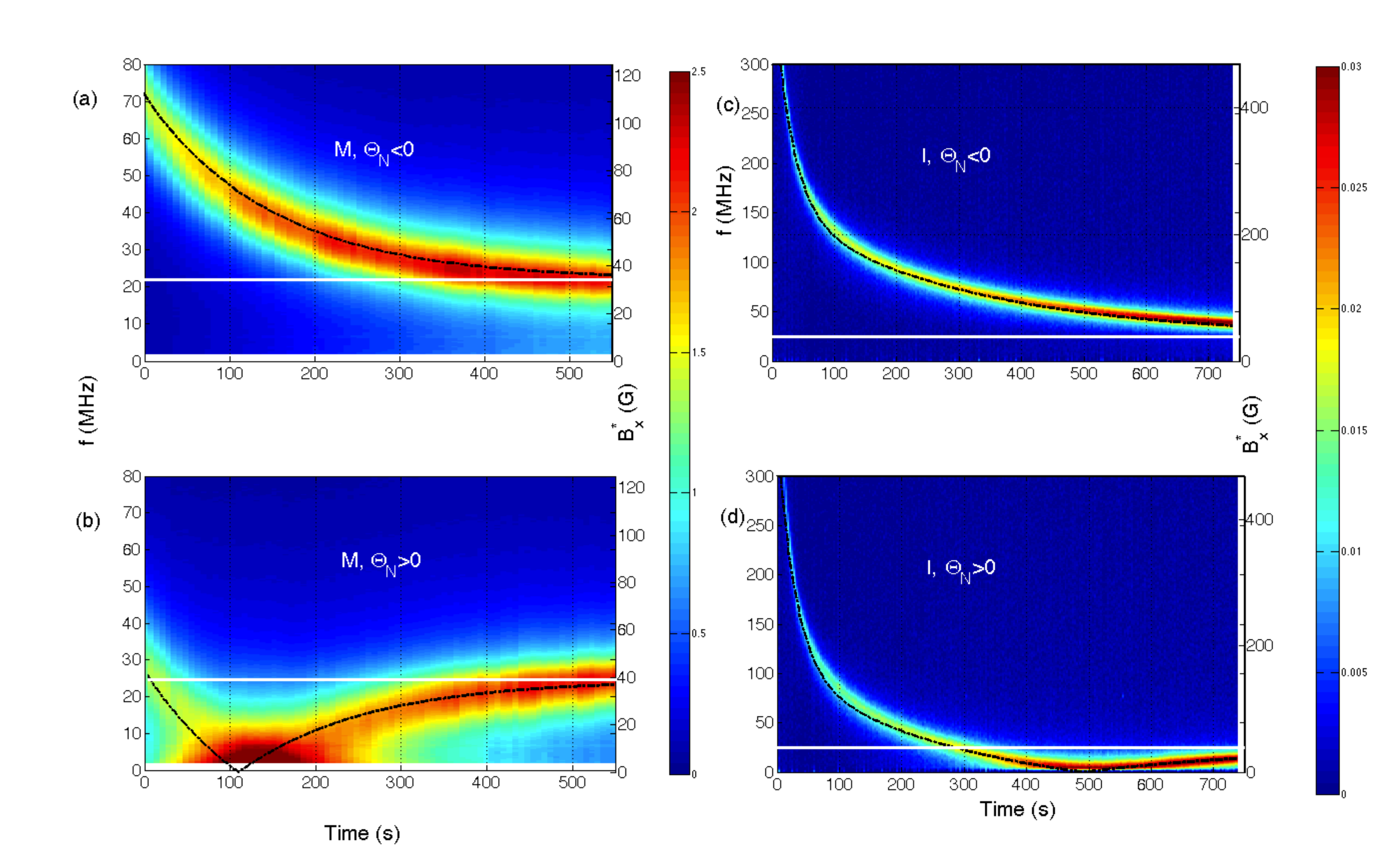} 
\caption{Electron spin noise spectra measured with time steps on the order of several seconds at the conducting (M) [panels (a) and (b)] and insulating (I) [(c), (d)] bulk $n$-type GaAs samples in the external magnetic field $B_{\rm ext}\approx  4$~mT. Panels (a), (c) correspond to a negative nuclear temperature ($\Theta_N<0$), where the nuclear field is parallel to the external one, and panels (b), (d) correspond to a positive temperature, where the nuclear and external fields are antiparallel. Black lines are the fitting results. White lines show the spin precession frequency of electrons in the external field. From Ref.~\cite{ryzhov15}. }\label{fig:overhauser}
\end{figure*}

\subsection{Electron spin polarization fluctuations under optical orientation conditions}\label{sec:optor}

The optical orientation method is a powerful tool to create non-equilibrium spin distributions in semiconductors: The spin-orbit coupling results in the transfer of the angular moment of absorbed photons to the electron spin system and generation of spin-polarized charge carriers~\cite{opt_or_book}. The problem of electron spin fluctuations investigation in a bulk semiconductor under the absorption of the circularly polarized radiation has been formulated and solved in Ref.~\cite{ivchenko73fluct_eng}. The detailed theory developed within the Greens function method included also the generation-recombination noise effects in the system of spin-polarized electrons. According to Ref.~\cite{ivchenko73fluct_eng} for the circularly polarized light propagating along $z$-axis the spectral density of spin $z$ component fluctuations, $\delta S_z = \sum_{\bm k} \delta s_{\bm k,z}$, for the ensemble of $N$ electrons takes a simple form
\begin{equation}
\label{sns:ivchenko}
(\delta S_z^2)_\omega = \frac{N(1-r_\omega P_s^2)}{2} \frac{T_s}{1+(\omega T_s)^2}.
\end{equation} 
Here $T_s$ is the spin lifetime which accounts for both the spin relaxation processes and the recombination of polarized electrons with unpolarized holes, $P_s$ is the steady-state degree of electron spin polarization, $r_\omega$ is a factor taking into account generation-recombination processes. Equation~\eqref{sns:ivchenko} demonstrates the suppression of electron spin fluctuations due to the spin-ordering. Analogous effect takes place in the static magnetic field $\bm B_{\rm ext} \parallel z$ as well, where thermal orientation of electron spins is realized. In such a situation, the factor describing spin fluctuations suppression takes a simple form $1-P_s^2$. This result can be derived from general arguments taking into account that, according to Eq.~\eqref{fourier},  $\int d\omega  (\delta S_z^2)_\omega = 2\pi \langle \delta S_z^2(t)\rangle$~\cite{ivchenko73fluct_eng}.

In addition to the electron spin orientation, the absorption of the circularly polarized light results in the ordering of crystal lattice nuclei spins~\cite{abragam,opt_or_book}. In typical experimental situations the nuclear spin system can be described by an effective temperature $\Theta_N$. The sign of the nuclear spin temperature $\Theta_N$ can be arbitrary, it is determined by the conditions of spin pumping. However, as a rule, the absolute value of nuclear spin temperature is substantially smaller than the lattice temperature~\cite{opt_or_book}. An external magnetic field induces nuclear spin polarization, which manifests itself as an effective magnetic field -- Overhauser field -- acting on electron spins. It has been shown in Ref.~\cite{PhysRevB.91.205301} that such a non-equilibrium spin polarization and, correspondingly, the Overhauser field, strongly modifies the spin noise spectrum of charge carriers. Particularly, if the external field $\bm B_{\rm ext}$ is applied perpendicularly to the probe light propagation direction $z$, the observed spin noise spectrum reads~\cite{gi2012noise,braun2007,PhysRevB.91.205301}
\begin{equation}
\label{sns:B}
(\delta S_z^2)_\omega = \frac{N}{4} \left(\frac{T_s}{1+(\omega-\Omega)^2 T_s^2} + \frac{T_s}{1+(\omega+\Omega)^2 T_s^2}\right),
\end{equation} 
where $\Omega = g \mu_B B_x^*/\hbar$ is the spin precession frequency in the total field  $B_x^*=B_{\rm ext} + B_N$, $\bm B_{N} \propto \alpha \bm B_{\rm ext}/\Theta_N$ is the Overhauser field, $g$ is the electron $g$-factor, $\alpha$ is the hyperfine coupling constant, the fluctuations suppression factor is omitted in Eq.~\eqref{sns:B}, nuclear spin fluctuations are neglected. It is seen from Eq.~\eqref{sns:B} that the peak in the spin noise spectrum at $\omega>0$ corresponds to the precession frequency $\Omega$, its position is controlled both by the external field and nuclear spin temperature.  

Nuclear spin relaxation is, as a rule, much slower as compared with the electron spin relaxation, hence, by measuring electron spin fluctuation spectra at different moments of time one can observe the dynamics of the Overhauser field $\bm B_N$, and, correspondingly, of the nuclear spin temperature $\Theta_N$. Such experiments have been carried out in Refs.~\cite{ryzhov15,2015arXiv150804968R} on $n$-type bulk GaAs samples placed in microcavities. Figure~\ref{fig:overhauser} presents the results of the  electron spin noise measurements on two samples with different doping level and, hence, different spin dynamics times of electrons and nuclei. Depending on the conditions of the nuclear spin system preparation, the nuclear spin temperature was either negative or positive, which corresponds to the Overhauser field being, respectively, parallel or antiparallel to the external one. Moreover, under the conditions of experiments in Ref.~\cite{ryzhov15} the absolute value of $B_N$ exceeded the external field. Therefore, for positive nuclear spin temperature, the frequency $\Omega$ of the electron spin precession goes through $0$ in the course of nuclear spin relaxation where the Overhauser field exactly compensates the external one, see Fig.~\ref{fig:overhauser}(b,d). In addition to the Overhauser field an effective ``optical'' magnetic field was uncovered in Ref.~\cite{2015arXiv150804968R}, which acts on the electron spins in the presence of elliptically polarized electromagnetic wave, whose frequency lies in the nominal transparency region of the crystal. This field is most probably caused by the circular dynamic Stark effect (or dynamic Zeeman effect)~\cite{cohen-atom-photon}. Results of Refs.~\cite{ryzhov15,2015arXiv150804968R} demonstrate the possibility to study experimentally the dynamics of non-equilibrium nuclear spin subsystem by means of the spin noise spectroscopy technique. We note also, that the nuclear spin fluctuations under the close-to-equilibrium conditions are also observed in the spin noise spectroscopy method~\cite{2015arXiv150605370B}.

\subsection{Spin noise in presence of a static electric field}\label{sec:dc}

A static electric field induces a drift of conduction electrons brining the electron system out of equilibrium. Depending on the field strength, rates of momentum and energy relaxation, different regimes of electron transport can be realized~\cite{gantmakher87}. Spin-orbit interaction induces a coupling between the orbital motion of electrons and their spin dynamics, hence, the presence of an external electric field may cause a considerable effect on spin phenomena in semiconductor systems~\cite{dyakonov_book}. For example, the effective Hamiltonian of conduction band electrons in quantum wells contains linear in the spin and wavevector terms~\cite{bychkov84,dyakonov86}:
\begin{equation}
\label{Hso}
\mathcal H_{so} = \frac{\hbar}{2}(\bm \Omega_{\bm k} \cdot \bm \sigma), \quad \Omega_{\bm k, \alpha} = \sum_{\beta} \gamma_{\alpha\beta} k_\beta.
\end{equation}
Specific form of the pseudotensor $\gamma_{\alpha\beta}$ and magnitudes of its components are determined by the material, crystallographic orientation of the quantum well, parameters of its potential and external fields~\cite{ivchenko05a}. The presence of spin-dependent terms, Eq.~\eqref{Hso}, in the effective Hamiltonian is equivalent to the presence of an effective magnetic field, which depends on the magnitude and direction of the wavevector and acts on the electron spin. Hence, the electric field $\bm E$ applied in the quantum well plane results both in the electron drift with the characteristic average wavevector $\bm k_{\rm dr} = e\bm E\tau_p/\hbar$, where $\tau_p$ is the electron momentum relaxation time, and in the generation of the effective magnetic field 
\begin{equation}
\label{drift}
\bm B_{\rm dr} = \frac{\hbar\bm \Omega_{{\rm dr}}}{g\mu_B}, \quad \bm \Omega_{{\rm dr}} = \bm\Omega_{\bm k_{\rm dr}}.
\end{equation}
This magnetic field, along with the external one, determines the electron spin precession frequency and, correspondingly, the position of the peak in the electron spin noise spectrum, cf. Eq.~\eqref{sns:B}. Such a shift of the electron spin precession frequency has been observed in the optical orientation~\cite{kalevich:230} and spin resonance~\cite{wilamowski07} experiments.  The theory of electron spin fluctuations in the presence of a weak pulling field has been developed in Ref.~\cite{Li:2013fk}, where the external electric field induced shift of the peak in the spin noise spectrum was predicted. The theory of Ref.~\cite{Li:2013fk} opens the way to measure the electron gas spin splittings via spin fluctuations.

Note that the effective field Eq.~\eqref{drift} leads to the current-induced spin orientation~\cite{aronov89:eng,edelstein90,PhysRevLett.93.176601,silov04,Ganichev2006127}. It is fundamentally non-equilibrium effect~\cite{PhysRevB.84.115303}. Current induced spin orientation, just like the optical orientation of electron spins, should manifest itself as a suppression of the spin fluctuations amplitude. Although a detailed theory of spin noise is absent for this case, the estimates show that the suppression effect is small and amounts to fractions of percent.

\begin{figure}
\includegraphics[width=0.8\linewidth]{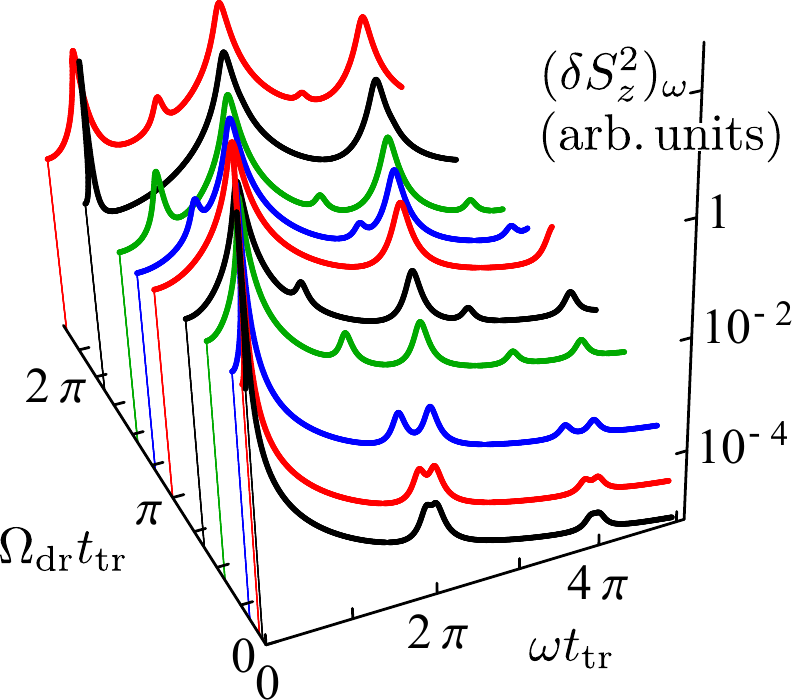} 
\caption{Spin noise spectra in the streaming regime, $\Omega_{\rm dr}$ is a mean frequency of electrons spin precession, $t_{tr}$ is the electron flight time until the optical phonon emission. From Ref.~\cite{PhysRevB.92.035437}. }\label{fig:stream}
\end{figure}

Spin noise of electron gas drastically changes in presence of moderately strong electric fields where the streaming regime of the electron transport is realized, and the electron distribution becomes needle-shaped~\cite{gantmakher87,ANDRONOV1992169}. In this regime, the electron accelerates ballistically until its energy reaches that of the optical phonon, the phonon emission takes place almost instantaneously and the electron returns to the zero-energy state, then its acceleration repeats. Periodic motion of electrons in the streaming regime gives rise to novel effects in spin dynamics~\cite{1367-2630-15-12-125003,Golub2014}. Spin fluctuations theory for this regime has been developed in Ref.~\cite{PhysRevB.92.035437}. The results of the spin noise spectra calculations for the streaming conditions are shown in Fig.~\ref{fig:stream}. It is demonstrated in Ref.~\cite{PhysRevB.92.035437} that the spin fluctuations spectrum consists of series of peaks with the frequencies
\begin{equation}
\label{comb:stream}
\omega_{nm} = \frac{2\pi n}{t_{tr}} + m \Omega_{\rm dr}, \quad n \in \mathbb Z, \quad m=-1,0,1,
\end{equation} 
where $\Omega_{\rm dr}$ is the average spin precession frequency in the needle-like distribution [cf. Eq.~\eqref{drift}], $t_{tr}$ is the time of electron acceleration until its energy being initially zero reaches the optical phonon energy. The intensities of the peaks depend on the electric field value, the spin-orbit coupling constants as well as on the electric field orientation with respect to the crystalline axes of the structure. The peak widths are related with the scattering time and spin relaxation times in the system. Reference~\cite{PhysRevB.92.035437} demonstrates the potential of the spin noise spectroscopy for studies of complex electron and spin dynamics in non-equilibrium conditions.

\subsection{Spin noise in presence of an alternating magnetic field}\label{sec:ac}

Non-stationary and non-equilibrium conditions in a spin system are brightly manifested in the presence of an alternating magnetic field.  Such a situation is well studied from the point of view of classical electron spin resonance, see review~\cite{Kalin:2006aa}. Spin fluctuations in the presence of alternating magnetic fields have been studied both theoretically, Refs.~\cite{braun2007,PhysRevB.91.155301}, and experimentally, Ref.~\cite{PhysRevLett.113.156601}. References~\cite{braun2007,PhysRevB.91.155301} predict the presence of harmonics in the spin noise spectrum with the frequencies being multiples of the alternating field frequency. For example, for linearly polarized alternating field $\bm B_{\rm ext}(t) = \bm B_0\cos{(\omega_{0} t + \varphi)}$, $\bm B_0 \parallel x$, following expression for the spin noise spectrum was derived in Ref.~\cite{braun2007}(see also \footnote{Notations of Ref.~\cite{braun2007} differ from those used here, the averaging over the phase $\varphi$ was also assumed in Ref.~\cite{braun2007}.})
\begin{equation}
\label{sns:ac}
( \delta S_z^2)_\omega =\frac{N}{4} \sum_{n=-\infty}^\infty J_n^2\left(\frac{g\mu_B B_0}{\hbar\omega_0}\right) \frac{T_s}{1+(\omega+n\omega_0)^2T_s^2},
\end{equation}
where $J_n(x)$ is the Bessel function of the order $n$. The presence of harmonics, $n\omega_0$, can be interpreted in terms of multiphoton processes by analogy with the classical result for ionization of an atom by a high-frequency field~\cite{keldysh_ion}: The quasi-stationary states of the system in the presence of an alternating magnetic field correspond to quasi-energies $\pm n\omega_0$, where $n$ is the number of the field quanta. Similar result was derived in Ref.~\cite{PhysRevB.91.155301} for spin fluctuations of localized charge carriers interacting both with the alternating field and with random nuclear fields.

Note, that due to $\bm k$-linear terms in the effective Hamiltonian~\eqref{Hso}, the electron spin noise spectrum similar to Eq.~\eqref{sns:ac} should be observed in the presence of an alternating electric field like the electric field induces, similarly to the magnetic field, the electron spin resonance~\cite{rashba64}.

Experimental studies of spin fluctuations in the presence of an alternating magnetic field were carried out in Ref.~\cite{PhysRevLett.113.156601} on an ensemble of $^{41}$K atoms. These experiments revealed, in addition to multiphoton processes, a number of interesting nonlinear effects, including Mollow triplet formation and dynamic Zeeman effect, and paved a way to study nonlinear optical effects by spin noise spectroscopy.

\section{Spin fluctuations in exciton systems}\label{sec:sns:excitons}

Optical excitation of semiconductors and semiconductor nanostructures with a quantum energy close to the band gap results in generation of excitons, electron-hole pairs bound by the Coulomb force. In doped quantum well and quantum dot structures three-particle complexes, trions or charged excitons, being a pair of same-sign charge carriers and an unpaired carrier with an opposite charge can be generated. Such complexes are fundamentally non-equilibrium, therefore, their spin fluctuations studies are rather involved and attract a considerable interest.

Below we present the results of theoretical and experimental studies of spin fluctuations in excitonic systems. We start with the structures with resident electrons where the light absorption and generation of charged excitons results in quantitative and qualitative changes of the resident electrons spin fluctuations. Afterwards we address the spin noise of excitons and exciton-polaritons in quantum microcavities.

\subsection{Effects of trion generation}\label{sec:trions}

\begin{figure}
\includegraphics[width=0.99\linewidth]{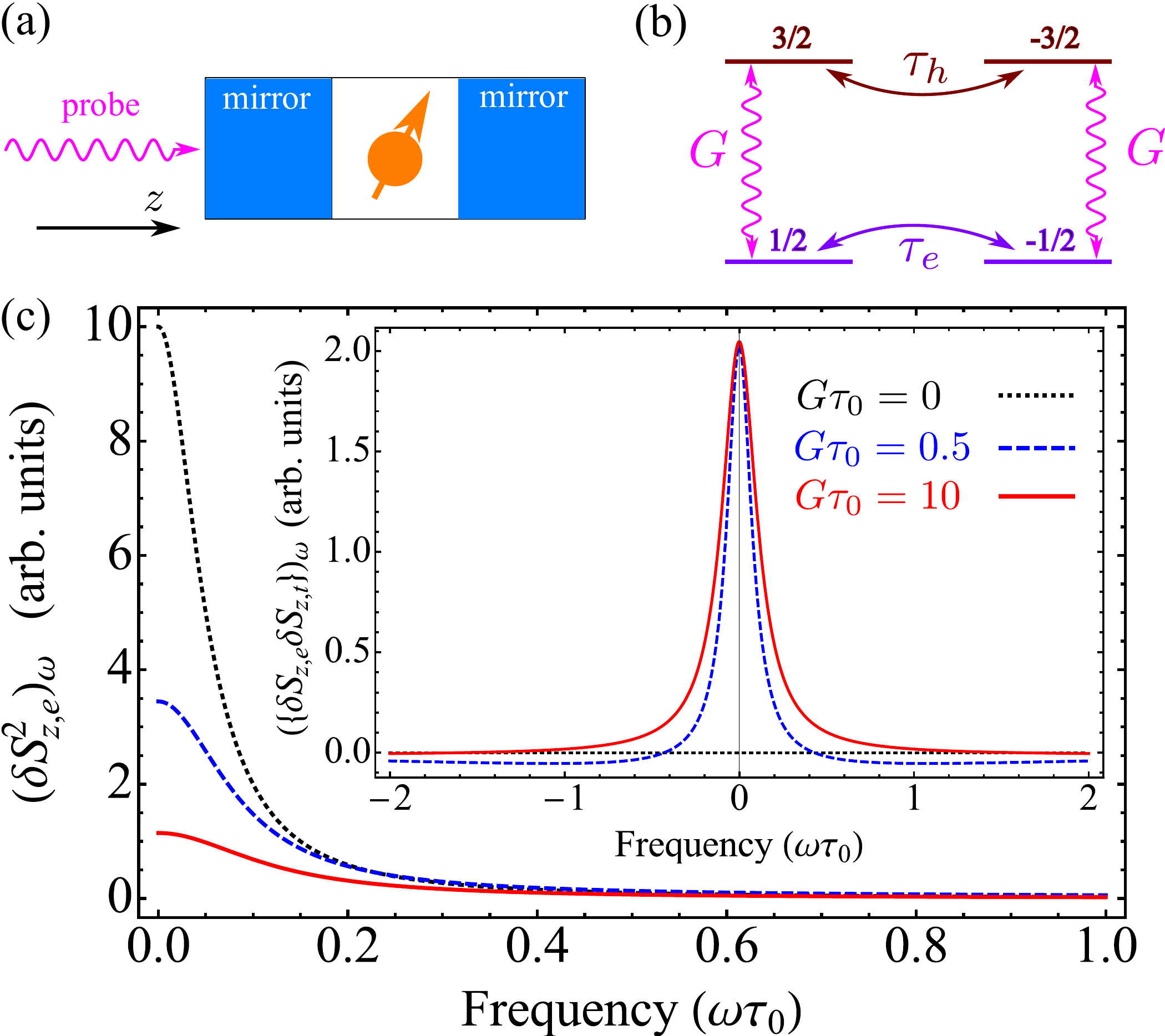} 
\caption{(a) Illustration of a quantum dot structure embedded into a microcavity. (b) Scheme of the states and transitions caused by the absorption of the probe beam and spin relaxation. (c) Electron spin noise spectra calculated for different trion generation rates $G\tau_0=0$, $G\tau_0 =0.5$ и $G\tau_0 = 10$, $\tau_e/\tau_0 = 20$, $\tau_h/\tau_0 = 5$. Inset shows the spectrum of correlation noise of the electron and trion.}\label{fig:QD}
\end{figure}

We start with the description of a simplest model illustrating the effect of the probe beam absorption and trion photogeneration on electron spin fluctuations. Consider a semiconductor quantum dot with a resident electron, whose ground state is two-fold spin degenerate, $s=\pm 1/2$. The excited state of the dot corresponding to a singlet trion with unpaired hole is also doubly-degenerate in the hole spin $j=\pm 3/2$, see Fig.~\ref{fig:QD}(a,b). Such a four-level model gained a considerable popularity for description of the charge carriers spin dynamics in single quantum dots and quantum dot ensembles~\cite{glazov:review,Arnold2015,PhysRevB.92.115305}. The coherent linearly polarized electromagnetic radiation with the frequency $\omega$ close to the trion photogeneration frequency $\omega_0$ is incident on the dot, the field is assumed to be strong enough to excite the trion. Such a situation can be also realized in  quantum microcavity structures operating in a weak coupling regime, see Fig.~\ref{fig:QD}(a) and Refs.~\cite{Arnold2015,PhysRevLett.112.156601}. It follows from the general equations for the single particle density matrix $\rho_{mm'}$, $m=\pm 1/2,\pm 3/2$, that the average occupancies of the electron, $f=(\rho_{1/2,1/2}+\rho_{-1/2,-1/2})/2$, and trion, $n=(\rho_{3/2,3/2}+\rho_{-3/2,-3/2})/2$, states obey a system of linear equations whose solutions have the form:
\begin{equation}
\label{fn:steady}
f=\frac{1}{2} \frac{1+G\tau_0}{1+2G\tau_0}, \quad n=\frac{1}{2}-f.
\end{equation} 
The trion generation rate in Eq.~\eqref{fn:steady}
\[
G=\frac{\gamma |V|^2}{(\omega - \omega_0)^2 + \gamma^2},
\]
$V = dE/(\sqrt{2}\hbar)$, $d$ is the matrix element of a dipole transition between the electron and the trion states, $E$ is the amplitude of the incident field, $\gamma$ is the damping rate of the off-diagonal elements of the density matrix, $\rho_{\pm 1/2, \pm 3/2}$, $\tau_0$ is the trion lifetime. One can also check that under the steady-state conditions the average $z$-components of electron, $S_{z,e}=(\rho_{1/2,1/2}-\rho_{-1/2,-1/2})/2$, and trion, $S_{z,t}=(\rho_{3/2,3/2}-\rho_{-3/2,-3/2})/2$, spins are zero. The density matrix method allows us to derive equations for the slow dynamics of spin correlators~\footnote{We assume hereafter that all characteristic rates of relaxation and generation are small as compared with $[(\omega - \omega_0)^2 + \gamma^2]^{1/2}$. The general case will be reported elsewhere.} 
\begin{equation}
\label{spin:corr}
\dot{\mathcal M} + \mathcal R \mathcal M=0.
\end{equation}
Here we introduced the correlators matrix $\mathcal M$ with the elements $\mathcal M_{ij}(\tau) =\langle \delta S_{z,i}(t+\tau)\delta S_{z,j}(t)\rangle$, where the subscripts $i,j=e,t$ enumerate the electron and trion states, and the matrix
\begin{equation}
\label{M:corr}
 \quad \mathcal R = \begin{pmatrix}
\frac{1}{\tau_e} + G & -\frac{1}{\tau_0} - G\\
-G & \frac{1}{\tau_h} + \frac{1}{\tau_0} + G
\end{pmatrix},
\end{equation}
describing the relaxation and generation processes, dot on top denotes the derivative over $\tau$. Two more parameters of the theory are introduced in Eq.~\eqref{M:corr}: the phenomenological spin relaxation times of the electron $\tau_e$ and of the hole-in-trion $\tau_h$, Fig.~\ref{fig:QD}(b). Equation~\eqref{spin:corr} is analogous to the general kinetic equation \eqref{kinetic:2} for the correlation function. According to the general theory, Eq.~\eqref{spin:corr} should be supplemented by initial conditions, the values of correlators matrix $\mathcal M(\tau=0)$. Direct calculation shows that $\langle \delta S_{z,e}^2\rangle = f/2$, $\langle \delta S_{z,h}^2 \rangle = n/2$, while single-time cross-correlators are zero. We introduce an auxiliary $2\times 2$ matrix, $\chi(\omega)$ according to 
\begin{equation}
\label{chi}
(-\mathrm i \omega + \mathcal R) \chi = \mathcal M(0),
\end{equation}
hence, the spin noise spectrum can be recast in agreement with Eq.~\eqref{fourier} and Refs.~\cite{ll10_eng,PhysRevB.90.085303} as
\begin{equation}
\label{sns:QD}
(\delta S_{z,i} \delta S_{z,j})_\omega = \frac{1}{2}[\chi_{ij}(\omega) + \chi_{ij}^*(-\omega) + \chi_{ji}(-\omega) + \chi_{ji}^*(\omega)].
\end{equation}
In particular, the electron spin noise can be written as
\begin{multline}
\label{sns:QD:1}
(\delta S_{z,e}^2)_\omega = \\
\frac{f}{2} \left[\tau_e^{-1} + G -\mathrm i \omega  - \frac{G(G+\tau_0^{-1})}{\tau_h^{-1} + \tau_0^{-1} + G -\mathrm i \omega } \right]^{-1} + {\rm c.c.}
\end{multline}
Similar expression can be derived for the trion spin fluctuations spectrum, $(\delta S_{z,t}^2)_\omega$. Note that the optical transitions between the electron and trion states at $G\ne 0$ result in a cross-correlation of the electron and trion spins described by the correlation function $\langle \{\delta S_{z,e}(t+\tau) \delta S_{z,t}(t)\}\rangle$; we remind that the curly brackets stand for the symmetrized product $\{AB\} = (AB+BA)/2$.

Typical electron spin noise spectra as well as spectra of cross-correlation electron and trion fluctuations are shown in Fig.~\ref{fig:QD}(c). It is seen that an increase in the probe beam power results in a reduction of the electron spin fluctuations amplitude due to population of the trion state and in a broadening of a peak due to the enhancement of the electron spin relaxation (for the parameters in question $\tau_e > \tau_h$). At typical conditions where the trion lifetime $\tau_0$ is considerably shorter as compared with those of electron and hole spin relaxation, the trion spin noise spectrum has a more complex shape and is characterized by the components of different widths ($\sim \tau_e^{-1}$, $\sim \tau_h^{-1} + \tau_0^{-1}$). 
An inset in Fig.~\ref{fig:QD}(c) shows the cross-correlation of electron and trion noise spectra, $(\{\delta S_{z,e} \delta S_{z,h}\})_\omega$. Interestingly, since for $\tau=0$ the cross-correlator $\langle \{\delta S_{z,e}(t+\tau) \delta S_{z,t}(t)\}\rangle$ turns to zero, the frequency integral $\int (\{\delta S_{z,e} \delta S_{z,h}\})_\omega=0$, therefore, the cross-correlation fluctuations spectrum changes its sign at a certain frequency. The sign change point depends on the probe power and relaxation times in the system.

\begin{figure}
\includegraphics[width=0.8\linewidth]{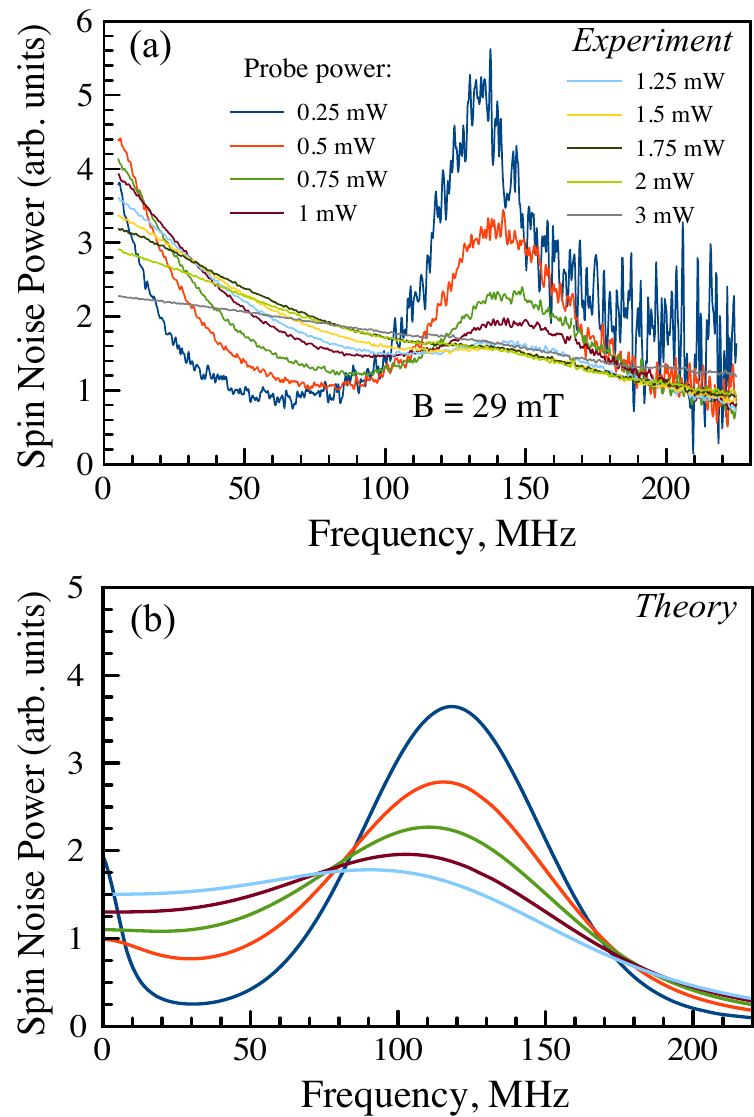} 
\caption{(a) Electron spin noise spectra in quantum well embedded into a microcavity measured at a fixed  $B=29$~mT, temperature $T =3.6$~K and different probe beam powers. Spectra are normalized at the probe power. (b) Calculated spin noise spectra of quantum well electrons at a fixed magnetic field $B=24$~mT and different trion generation rates $G = 0\ldots 5\times 10^{8}$~s$^{-1}$ (in equal steps in $G$). From Ref.~\cite{PhysRevB.89.081304}. }\label{fig:trions}
\end{figure}

Spin fluctuations of a single charge carrier in a quantum dot have been experimentally studied in Ref.~\cite{PhysRevLett.112.156601}. The effects of the trion generation on the electron spin noise have been experimentally and theoretically studied in Ref.~\cite{PhysRevB.89.081304} for a quantum well microcavity structure. The microcavity operated in the strong-coupling regime, both exciton and trion polaritons have been observed in the reflection spectrum.  The results of the resident electrons spin noise spectra measurements presented in Fig.~\ref{fig:trions}(a) demonstrate a drastic effect of the probe beam absorption on the electron spin noise. In particular, an increase of the power at a fixed transverse magnetic field results in the broadening of the noise spectrum as well as in almost complete suppression of the precession peak. These experimental data are described by the model developed in Ref.~\cite{PhysRevB.89.081304}, which takes into account both non-resonant excitation of trions and electron spin precession in the external field and in the field of nuclear spin fluctuations. The suppression of the precession peak with an increase of the probe beam power is caused, mainly, with the induced electron spin relaxation anisotropy due to its ``pumping'' through the trion state, as well as with absence of the electron spin precession in the trion, since the transverse field practically does not result in a spin splitting of heavy hole states~\cite{Mar99}. The results of calculations presented in Fig.~\ref{fig:trions}(b) describe satisfactorily the experimental data. Differences in the zero-frequency peak amplitude can be related with the neglected contributions of the photoexcited excitons and trions to the observed Kerr rotation noise spectrum as well as with possible non-linear processes in this system~\cite{ryzhov:jap}.

\subsection{Exciton spin fluctuations under nonresonant pumping}\label{sec:excitons}

Light absorption in undoped semiconductors and semiconductor nanostructures results in the formation of neutral excitons. In absence of an external pumping there are no excitons, therefore, exciton spin noise arises due to the pumping only. The theory of exciton and exciton-polariton spin noise has been developed in Refs.~\cite{PhysRevB.90.085303,glazov_sns_pol,PhysRevB.91.161307}. 

Heavy-hole exciton spin states in quantum well structures based on GaAs-like semiconductors are characterized by the growth axis $z$ component, $m_z = S_z + J_z$, of the total spin of the electron, $S_{z}=\pm 1/2$, and the hole, $J_z = \pm 3/2$. The exchange interaction between the electron and the hole splits the quadruplet of excitonic states into two doublets, with $m_z = \pm 1$ (radiative, optically active or ``bright'' doublet) and $m_z = \pm 2$ (``dark'' doublet)~\cite{ivchenko05a}. A transverse magnetic field $\bm B \perp z$ mixes ``bright'' and ``dark'' excitonic states. Exciton spin dynamics is mainly governed by an interplay of the Zeeman effect of the external magnetic field and the exchange interaction between the electron and the hole~\cite{toulouse1}. The role of such a competition in excitonic spin fluctuations has been theoretically investigated in Ref.~\cite{PhysRevB.90.085303}. 

It follows from Ref.~\cite{PhysRevB.90.085303} that if the exchange splitting between the optically active and inactive doublets $\delta_0$ is sufficiently large as compared with the decay rates of excitons and spins expressed in energy units, then in the exciton spin noise spectra at small and moderate magnetic fields, $B\lesssim |\delta_0/(g \mu_B)|$ (recall that $g$ is the electron $g$-factor, hole Zeeman effect in the transverse field is negligibly small~\cite{Mar99}), there is one peak centered at the zero frequency and corresponding to correlated fluctuations of the electron and hole spins. An increase in the magnetic field results in the broadening and suppression of the peak. In large magnetic fields, $B\gtrsim |\delta_0/(g \mu_B)|$, the peak at the combination frequency
\begin{equation}
\label{comb}
\Omega'=  \sqrt{(g\mu_B B)^2 + \delta_0^2}/\hbar,
\end{equation}
appears in the spin noise spectrum~\cite{PhysRevB.90.085303}, which corresponds to the electron spin precession in the effective field being the sum of the external and exchange ones. The peak widths in the spin noise spectra are controlled by the lifetimes of bright and dark excitonic states, rates of the charge carriers spin relaxation and the exciton generation rate.

In the opposite limit where the exchange interaction is small compared with the exciton levels broadening, it plays almost no role in spin fluctuations. The calculations of Ref.~\cite{PhysRevB.90.085303} show that the exciton spin noise spectrum in such a case comprises two peaks corresponding to independent fluctuations of the electron and hole spins. The sensitivity of the spin fluctuation spectra of excitons to the occupation numbers of bright and dark states opens up a possibility to study the dark exciton spin dynamics by means of the spin noise spectroscopy.

 \subsection{Spin fluctuations of exciton-polaritons}\label{sec:polaritons} 

Excitons -- the electron-hole pairs -- possess integer spin and can demonstrate effects inherent to bosons~\cite{Moskalenko62:eng,suris_Opt_prop,keldysh68a}. Theory of excitonic spin fluctuations proposed in Ref.~\cite{PhysRevB.90.085303} is limited by weak pumping conditions where the occupancy of the exciton state is small as compared with unity. In this case, the effects of quantum statistics of excitons are unimportant. Qualitatively different situation can be realized in quantum microcavities structures in the strong coupling regime. In these systems one can achieve a macroscopic occupancy of exciton-polariton states and these quasiparticles demonstrate prominent bosonic effects~\cite{microcavities,sanvitto_timofeev}. Noises in polaritonic systems have been theoretically studied in Refs.~\cite{Keld_Tikh,Gippius86}.

\begin{figure}
\includegraphics[width=0.8\linewidth]{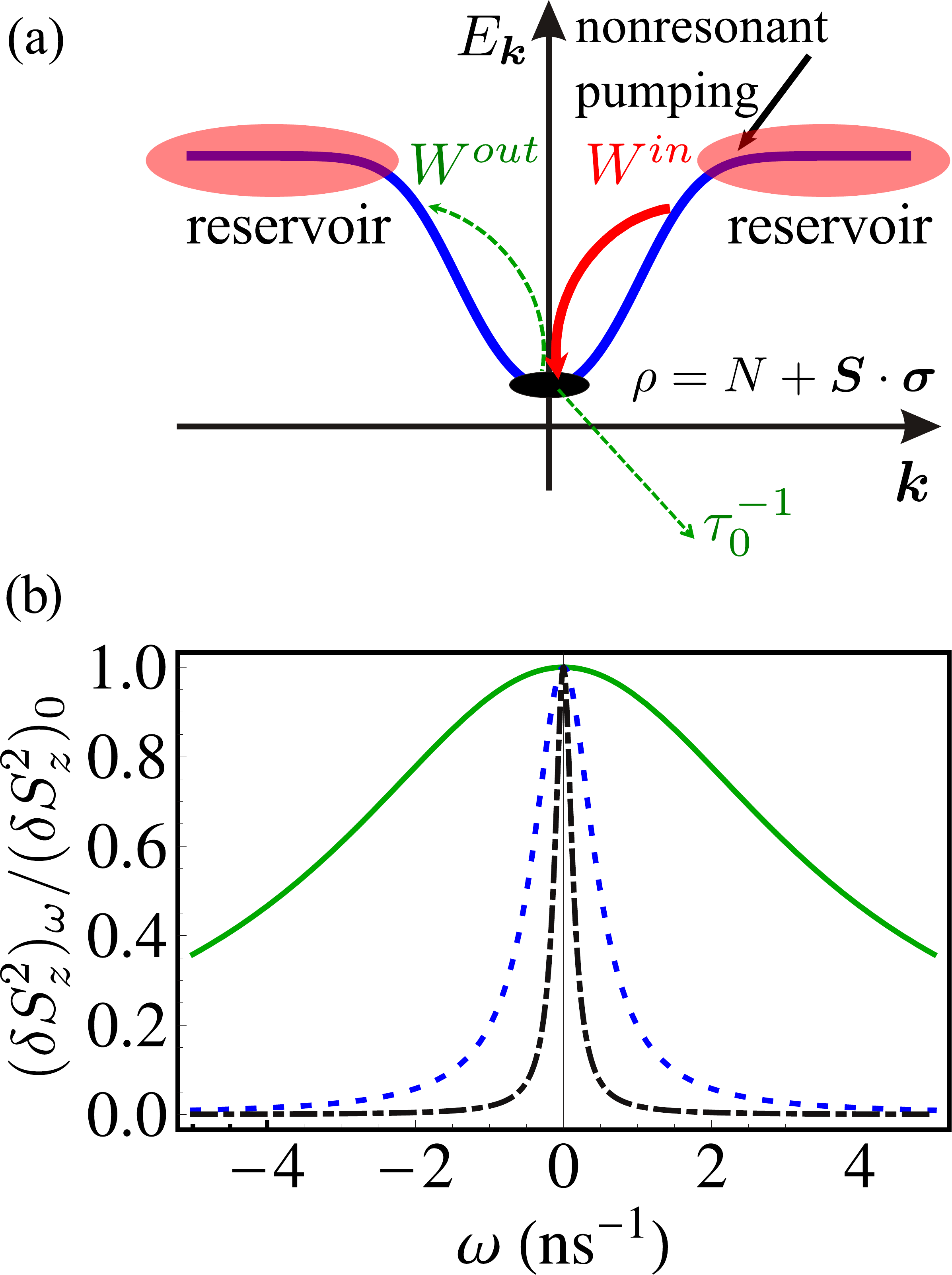} 
\caption{(a) Illustrative scheme of a microcavity structure pumping. The reservoir of particles formed at a non-resonant excitons is shown as well as the ground state, whose spin fluctuations are studied. $W^{in}$ and $W^{out}$ are the transition rates of polaritons to and from the ground state, respectively, $\tau_0$ is the polariton lifetime in the ground state.  (b) Spin noise spectra calculated for different number of particles in the ground state  $\langle N \rangle = 10$ [green solid curve], 100 [blue dotted], и 1000 [black dash-dotted]. The parameters of calculation are $\tau_0 = 25$~ps, $W_{out} = 0$, and $\tau_s = 10$~ns.  From Ref.~\cite{glazov_sns_pol}. }\label{fig:polaritons}
\end{figure}

Effects of macroscopic occupancy on spin fluctuations of exciton-polaritons have been theoretically analyzed in Ref.~\cite{glazov_sns_pol}. The simplest model was accepted accounting for the formation of exciton-polariton reservoir under non-resonant pumping and polariton relaxation from the reservoir towards the ground state due to their interactions with phonons and exciton-polaritons, Fig.~\ref{fig:polaritons}(a). It is sufficient to limit the consideration of exciton-polaritons by bright excitonic spin doublet with $m_z = \pm 1$, since ``dark'' excitons do not couple to light and do not form polaritons. Therefore, the treatment of exciton-polariton spin dynamics and fluctuations is quite similar to that developed in Sec.~\ref{sec:formalism} for electrons. In fact, exciton-polariton spin density matrix is a $2\times 2$ matrix and can be parametrized by spin-average occupancy of the state, $N$, and the pseudospin vector, $\bm S$, whose components characterize the degrees of circular and linear polarizations of polaritons.  The conclusion that the accumulation of polaritons in  the ground state results in a strong slow-down of spin fluctuations and, correspondingly, to the narrowing of the line in the spin noise spectrum is the main result of Ref.~\cite{glazov_sns_pol}. It is illustrated in Fig.~\ref{fig:polaritons}(b), where the calculated spin noise spectra of exciton-polaritons are shown for different mean occupancies of the ground state.

The physical reason of such a slow-down of temporal fluctuations and narrowing of the noise spectrum is related with the Bose-stimulation of the particle scattering processes. One can readily convince oneself of this fact considering the collision integral of polaritons with phonons, which describes the incoming processes to the ground state from the reservoir and the outcoming processes from the ground state to the reservoir. The collision integral describing the rate of the ground state population variation has a form~\cite{glazov_sns_pol}
\begin{equation}
\label{Q:pol}
Q_n\{N\} = -N/\tau_0 - W^{out} N + W^{in} (1+N),
\end{equation}
where $\tau_0$ is the polariton lifetime in the ground state caused by the trasnparency of the mirrors, $W^{in}$ and $W^{out}$ are the transition rates of polaritons to the ground state from the reservoir and back, respectively. For the steady state with the ground state occupancy $\langle N\rangle$, the collision integral vanishes, hence, $\langle N \rangle = W^{in}\tau_0/[1+(W^{out} - W^{in}) \tau_0]$. Relaxation of small fluctuations of occupation number $\delta N = N - \langle N \rangle$ and spin $\delta \bm S$ is also described by the collision integral Eq.~\eqref{Q:pol} and characterized by the correlation time $\tau_c$:
\begin{equation}
\label{tau_c:pol}
\frac{1}{\tau_c} = \frac{\tau_0^{-1} + W^{out}}{1+\langle N \rangle},
\end{equation}
which is the longer, the larger the ground state occupancy. In fact, the fluctuations are supported by the stimulated income of the polaritons to the ground state from the reservoir.
 
A question about the mean square of the spin fluctuation of polaritons has been also studied in Ref.~\cite{glazov_sns_pol}. The special complexity of this problem is related with the necessity to determine the full distribution function of the system of bosons. In fact, the ground state statistics, $P(N)$, i.e. the probability to find  $N$ particles in this state, depends strongly on the pumping mechanisms of the system and polariton-polariton interactions~\cite{sanvitto_timofeev}. Therefore, the mean square of the particle number or spin fluctuation for bosons cannot be expressed via the occupancy of the state only [cf. Eq.~\eqref{same:t}], it is determined by one more parameter of the system: the second-order coherence $g^{(2)}$~\cite{glazov_sns_pol}:
\begin{equation}
\label{sns:ampl}
\langle \delta N^2\rangle = \langle N\rangle [1 + (g^{(2)} - 1)\langle N\rangle], \quad \langle  \delta S_z^2\rangle = \langle \delta N^2\rangle/2.
\end{equation}
Note that the expression for the mean square of the spin fluctuation [second equality in Eq.~\eqref{sns:ampl}] holds in absence of the spin polarization and interactions of polaritons. It was shown in Ref.~\cite{glazov_sns_pol} that due to interparticle interactions the spin noise spectrum can be also quite sensitive to the particle statistics. Note, that interaction between the particles results also in non-trivial features in space-time correlations of exciton-polaritons and can provide propagation of spin waves in such a system~\cite{PhysRevB.91.161307}.

\section{Conclusion and prospects}\label{sec:concl}

Spin noise spectroscopy is rapidly developing and progressively taking an important place in the arsenal of experimental methods to study the spin dynamics of the charge carriers and the charge carrier complexes in semiconductors. To date, a number of convincing experimental evidences and detailed theoretical models of spin fluctuations in non-equilibrium conditions are accumulated. Nevertheless, in our opinion, the studies in this area are far from being complete. Particularly, a problem of the source of simultaneous spin correlations of electrons caused by the interparticle interaction requires further investigations. The problem of space-time spin fluctuations in electron and exciton systems seems interesting as well (equilibrium case was considered in Refs.~\cite{PhysRevB.91.161307,PhysRevB.92.045308}). The problems of spin fluctuations of single electrons and excitons in structures with strong light-matter coupling: quantum dot microcavities operating in the strong coupling regime are, beyond doubt, of current importance. Certainly, experimental studies of such kind in presence of strong static or alternating electromagnetic fields would make it possible to confirm developed models of non-equilibrium electron spin noise and would put new questions requiring theoretical analysis.

\acknowledgments
Authors is grateful to L.E. Golub, E.L. Ivchenko, D.S. Smirnov, and V.S. Zapasskii for valuable discussions. The work was partially supported by Russian Science Foundation, grant 14-12-01067, RF President grant MD-5726.2015.2 and Dynasty Foundation.

    

\end{document}